# Quantum circuit to estimate pi using quantum amplitude estimation

Takuma Noto

*Institute of Technology, Shimizu Corporation, 3-4-17, Etchujima, Koto-ku, Tokyo 135-8530, Japan*

*London Centre for Nanotechnology, University College London, 17-19 Gordon Street, London WC1H 0AH, United Kingdom*

Email: t.noto@shimz.co.jp

October 21, 2020

**Abstract**

This study presents a quantum circuit for estimating the pi value using arithmetic circuits and by quantum amplitude estimation. We review two types of quantum multipliers and propose quantum squaring circuits based on the multiplier as basic arithmetic circuits required for performing quantum computations. The squarer realized by a quantum adder with the gate size of $O(n)$ requires $O(n^2)$ gates and at least one ancillary qubits, while that realized by using quantum Fourier transform (QFT) requires $O(n^3)$ gates without ancillary qubit. The proposed quantum circuit to estimate pi is based on the Monte Carlo method, quantum amplitude estimation, and quantum squarer. By applying the quantum squarer using QFT, the circuit was implemented in $4n + 1$ qubits at $2^{2n}$ sampling. The proposed method was demonstrated using a quantum computer simulator with $n$ being varied from 2 to 6, and the obtained result was compared with the one obtained by performing a classical calculation.

## 1. Introduction

Quantum computers are expected to perform high-speed calculations in many fields. Recently, practical algorithms for Noisy Intermediate-Scale Quantum (NISQ) computer based on the use of fewer qubits and gates than the ideal ones for the future have been developed [1-3]. The Monte Carlo method on quantum computer by applying quantum amplitude estimation [4, 5] is one of the algorithms reported to have less computational complexity than the classical methods [6]. Further, [5] demonstrated a quantum circuit for Monte Carlo integration of the sine function using a newly proposed quantum amplitude estimation method with few controlled gates and no auxiliary qubits for NISQ. Financial problems have been studied as one of the applications of the Monte Carlo method on quantum computers [7], and classical simulations in various fields can be expected to be faster. This paper implements a pi estimation circuit as an example of the applications of the Monte Carlo method on quantum computers. Note that quantum algorithm for calculating pi without the Monte Carlo method can be found in [8, 9].

Further, various small arithmetic operation circuits such as adder and multiplier (multiplier accumulator) are necessary when building a large circuit to solve a complex problem. An efficient adder works as follows: $|x\rangle|y\rangle \to |x\rangle|x + y\rangle$, where $|x\rangle$, $|y\rangle$, and $|x + y\rangle$ are quantum registers with representative states of integer numbers $x$, $y$, and $x + y$, respectively. This operation changes the state of the second register into $|x + y\rangle$ even if $|x\rangle$ and $|y\rangle$ are a superposition of arbitrary integers.

Various quantum adders based on a classical algorithm such as ripple-carry adder, carry look-ahead adder, or a combination of these algorithms have been implemented in quantum circuits [10, 11]. For example, a ripple-carry adder for two $n$-bit binary numbers can be implemented in $O(n)$ gates without the use of an ancillary qubit [11]. As another approach, a quantum Fourier transform (QFT) based quantum adder can be implemented in $O(n^2)$ gates with no ancillary qubit [12]. Quantum algorithms of a multiplier for two quantum registers have also been proposed [13-17]. Particularly, the schoolbook method can be implemented by repeating the quantum adder [16]. On the other hand, a QFT adder based multiplier consists of $O(n^3)$ gates with no ancillary qubit [18-20].

This paper first reviews two types of quantum multipliers: a quantum multiplier by schoolbook method based on quantum adder circuit and the QFT multiplier. Second, this paper proposes two types of quantum squaring circuits based on the quantum multipliers. Finally, a quantum circuit to estimate pi by using the proposed quantum squarer and quantum amplitude estimation is demonstrated and the results obtained by the quantum circuit are compared with the results obtained by performing a classical calculation.

## 2. Arithmetic circuit

### 2.1 Quantum multiplier

By inputting first quantum register $|a\rangle_1 = |a_{n-1}\rangle_1 \otimes |a_{n-2}\rangle_1 \otimes \ldots \otimes |a_0\rangle_1$ with $n$-qubit representing integer $a = 2^{n-1}a_{n-1} + 2^{n-2}a_{n-2} + \ldots + 2^0 a_0$, second quantum register $|b\rangle_2 = |b_{m-1}\rangle_2 \otimes |b_{m-2}\rangle_2 \otimes \ldots \otimes |b_0\rangle_2$ with $m$-qubit representing integer $b = 2^{m-1}b_{m-1} + 2^{m-2}b_{m-2} + \ldots + 2^0 b_0$, and third quantum register $|c\rangle_3 = |c_{l-1}\rangle_3 \otimes |c_{l-2}\rangle_3 \otimes \ldots \otimes |c_0\rangle_3$ with $l$-qubit representing integer $c = 2^{l-1}c_{l-1} + 2^{l-2}c_{l-2} + \ldots + 2^0 c_0$, a multiplier changes the third register as follows:

$$|a\rangle_1 |b\rangle_2 |c\rangle_3 \rightarrow |a\rangle_1 |b\rangle_2 |c + ab \bmod 2^l\rangle_3, \qquad (1)$$

where the modulo in the third register on the right hand side implies that this operation may result in overflow when $c + ab$ is more than $2^l$. Normally, there is no overflow when $c = 0$ and $l \geq m + n$.

#### 2.1.1 Classical quantum multiplier

A schoolbook multiplication by applying an adder repeatedly is one of the classical algorithms for multiplying numbers [16]. Figure 1 shows a quantum adder on the first register $|x\rangle$ as addend and the second register $|y\rangle$ as augend, which changes the state of the second register into $|x + y\rangle$. The multiplier consists of controlled quantum adders by applying on $|a_s\rangle_1$ ($s = 0, 1, \ldots, n - 1$) as control, $|b\rangle_2$ as addend, and the upper $l - s$ digits of $|c\rangle_3$ as augend as shown in Fig. 2. By applying a series of controlled quantum adders, the value stored in the third register changes as follow.

$$\begin{aligned}
|c\rangle_3 &\rightarrow |c + 2^0 a_0 b \bmod 2^l\rangle_3 \\
&\rightarrow |c + 2^1 a_1 b + 2^0 a_0 b \bmod 2^l\rangle_3 \\
&\vdots \\
&\rightarrow |c + 2^{n-1} a_{n-1} b + \cdots + 2^1 a_1 b + 2^0 a_0 b \bmod 2^l\rangle_3 = |c + ab \bmod 2^l\rangle_3. \quad (2)
\end{aligned}$$

This multiplier circuit consists of $n$ controlled quantum adders and the ancillary qubits required by the quantum adder. For two $n$-qubit multiplications using a quantum adder whose gate size is $O(n)$ and no ancillary qubit, the multiplier can be implemented in $O(n^2)$ gate without ancillary qubits.

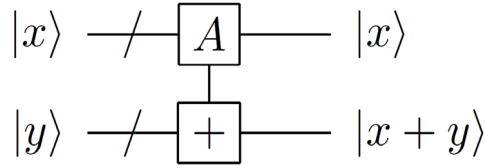

**Fig. 1** Quantum circuit of an adder—in the circuit, two gates represented by $A$ and $+$ act on first and second quantum register, respectively, to modify the state of the second register

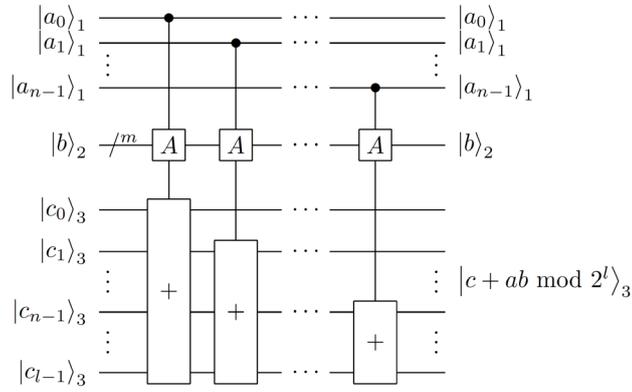

**Fig. 2** Quantum circuit of a quantum multiplier that uses quantum adders

*2.1.2 QFT multiplier*

QFT multiplier can be illustrated as shown in Fig. 3 with $O(n^3)$ gates and no ancillary qubits [20]. First, QFT is applied to the third register and its state changes from $|c\rangle_3 \to |\varphi(c)\rangle_3$, where $u$-th ($u = 0, 1, …, l − 1$) qubit of the third register changed by QFT can be expressed as follows:

$$|c_u\rangle_3 \to \frac{1}{\sqrt{2}}(|0\rangle_3 + e^{2\pi i 0.c_u c_{u-1} \cdots c_0}|1\rangle_3) = |\varphi_u(c)\rangle_3. \quad (3)$$

Then, the phase of the third register is rotated by applying controlled-$2^s \Sigma$ gate, which acts on $|a_s\rangle_1$ as control and on $|b\rangle_2$ and $|\varphi(c)\rangle_3$ as input. The circuit of the gate consists of a series of a controlled-controlled-$R_j$ gate that acts on $|a_s\rangle_1$ and $|b_t\rangle_2$ as control, and rotate the $u$-th qubit of the third register with $j = − s − t + u + 1$ as shown in Fig. 4, where $R_j$ gate can be expressed as follows:

$$R_j = \begin{pmatrix} 1 & 0 \\ 0 & e^{\frac{2\pi i}{2^j}} \end{pmatrix}. \tag{4}$$

Hence, the $R_j$ gate will be an identity operator when $j \leq 0$.

Now, non-controlled $2^0 \Sigma$ operations when $s = 0$ are equal to phase rotations of the QFT adder [12] that change the value in the third register as follows: $|\varphi(c)\rangle_3 \to |\varphi(c + b \mod 2^l)\rangle_3$. Controlled $2^s \Sigma$ operation with $|a_s\rangle_1$ as control, further multiplies the phase of $2^s a_s$ and the third register changes as follows: $|\varphi(c)\rangle_3 \to |\varphi(c + 2^s a_s b \mod 2^l)\rangle_3$. Thus, the third register changes in the following manner by repeating controlled-$2^s \Sigma$ gate for $s = 0, 1, \ldots, n-1$.

$$\begin{aligned}
|\varphi(c)\rangle_3 &\to |\varphi(c + 2^0 a_0 b \mod 2^l)\rangle_3 \\
&\to |\varphi(c + 2^1 a_1 b + 2^0 a_0 b \mod 2^l)\rangle_3 \\
&\vdots \\
&\to |\varphi(c + 2^{n-1} a_{n-1} b + \cdots + 2^1 a_1 b + 2^0 a_0 b \mod 2^l)\rangle_3 \\
&= |\varphi(c + ab \mod 2^l)\rangle_3.
\end{aligned} \tag{5}$$

Finally, the third register is transformed by using inversed-QFT to attain the state $|c + ab \mod 2^l\rangle_3$.

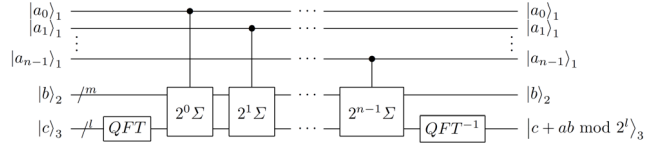

**Fig. 3** Quantum circuit of QFT multiplier consisting of a series of controlled-$2^s \Sigma$ gates for $s = 0, 1, \ldots, n-1$—in the circuit, *QFT* and *QFT*$^{-1}$ stand for the QFT circuit and its inverse circuit, respectively

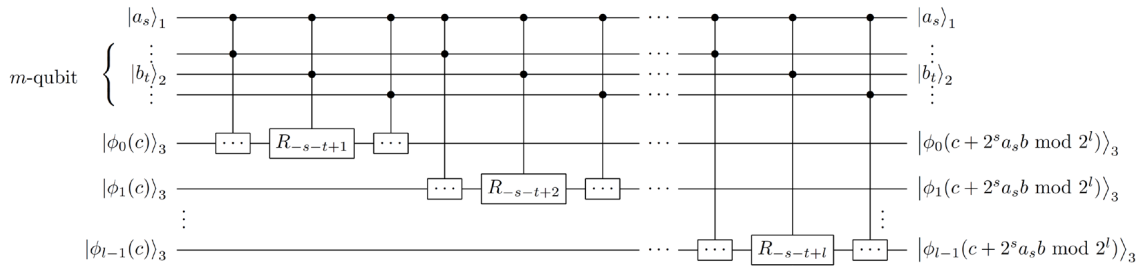

**Fig. 4** Quantum circuit of controlled-$2^s \Sigma$ gate.

## 2.2 Quantum squarer

A squaring circuit is used to calculate the squared number of integers in the superposition. The circuits can be implemented similarly as the multiplier, by omitting one of its inputs. Inputting the first quantum register $|a\rangle_1$ with $n$-qubit and the second quantum register $|b\rangle_2$ with $m$-qubit, a squarer circuit changes the second register as follows:

$$|a\rangle_1 |b\rangle_2 \rightarrow |a\rangle_1 |b + a^2 \bmod 2^m\rangle_2, \qquad (6)$$

where the modulo in the second register on the right hand side signifies that an overflow may occur when $b + a^2$ is more than $2^m$. Normally, there is no overflow when $b = 0$ and $m \geq 2n$.

### 2.2.1 Classical quantum squarer

The quantum squarer circuit shown in Fig.5 consists of a series of controlled quantum adder with an ancillary qubit $|0\rangle_A$. By applying controlled quantum adder on ancillary qubit as target after copying the state of $|a_s\rangle_1$ by CNOT gate, on $|a\rangle_1$ as addend, and on upper $m - s$ digits of the second register as augend, $2^s a_s a$ is added to the second register. After applying a series of controlled quantum gate, the state of the second register becomes $|b + a^2 \bmod 2^m\rangle_2$.

This quantum squarer consists of $n$ quantum adders with $1 + N_A$ ancillary qubits, where $N_A$ denotes the number of ancillary qubits of the quantum adder. The order of its gate size is equal to that of a quantum multiplier for two $n$-qubit binary numbers presented in section 2.1.1.

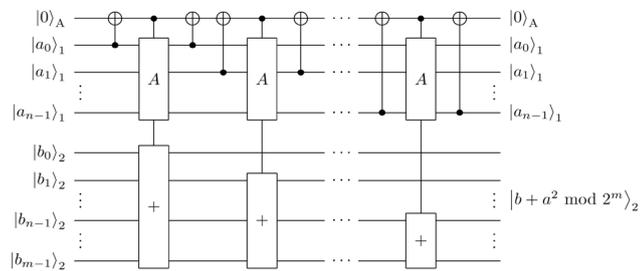

**Fig. 5** Quantum circuit of quantum squarer obtained using a quantum adder

*2.2.2 QFT squarer*

A QFT based quantum squarer is shown in Fig. 7. First, the second quantum register $|b\rangle_2$ is translated to $|\varphi(b)\rangle_2$ by QFT, where the *u*-th ($u = 0, 1, \ldots, m - 1$) qubit of the second register changes as follows:

$$|b_u\rangle_2 \to \frac{1}{\sqrt{2}}\left(|0\rangle_2 + e^{2\pi i 0.b_u b_{u-1} \ldots b_0}|1\rangle_2\right) = |\varphi_u(b)\rangle_2. \tag{7}$$

Then, a series of $2^s \Sigma_s$ gate inputting $|a\rangle_1$ and $|\varphi(b)\rangle_2$ rotate the second register by an encoded value of $2^s a_s a$. The *u*-th qubit of the second register is rotated by controlled-controlled-$R_{-s-t+u+1}$ gate (which is controlled by $|a_s\rangle_1$ and $|a_t\rangle_1$ ($t = 0, 1, \ldots, n - 1, t \neq s$)) and by controlled-$R_{-2s+u+1}$ gate (which is controlled by $|a_s\rangle_1$) as shown in Fig. 8. The $2^s \Sigma_{p,s}$ gate changes the second quantum register as follows: $|\varphi(b)\rangle_2 \to |\varphi(b + 2^s a_s a \bmod 2^m)\rangle_2$, and a series of the $2^s \Sigma_s$ gates change it as follows: $|\varphi(b)\rangle_2 \to |\varphi(b + a^2 \bmod 2^m)\rangle_2$.

This QFT squarer requires $O(n^3)$ gate size with no ancillary qubits as well as QFT multiplier for two *n*-qubit binary numbers

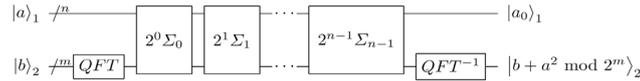

**Fig. 6** Quantum circuit of QFT squarer consisting of a series of $2^s \Sigma_s$ gates where $s = 0, 1, \ldots, n - 1$

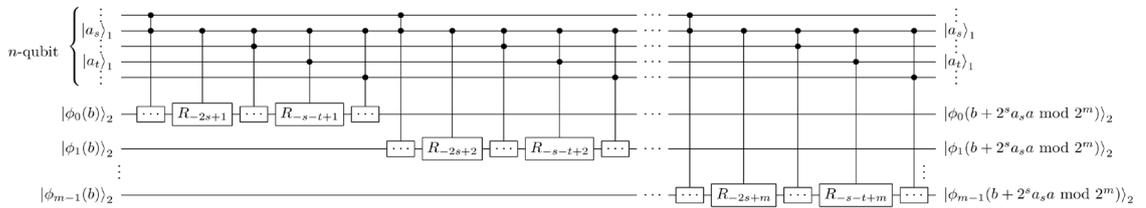

**Fig. 7** Quantum circuit of $2^s \Sigma_s$ gate

## 3. Quantum circuit to estimate pi

### 3.1 Preparation

The Monte Carlo method in classical computers solves a problem using a random number such as estimating the expected value of a problem by using repeated random sampling. The accuracy of the result obtained in such a task can be improved by increasing the computational complexity or number of samplings, which require $O(1/\epsilon^2)$ to produces an estimated error $\epsilon$.

In quantum computing, new algorithms have been proposed for evaluating the expected value by quantum amplitude estimation [6]. Suppose $f(x)$ is a classical real value function for $n$-bit binary input $x \in \{0, 1\}^n$ and $p(x)$ is its probability, the expected value of all possible numbers in $x$ can be evaluated as follows:

$$\mathbb{E}[f(x)] = \sum_{x=0}^{2^n-1} f(x)\,p(x). \tag{8}$$

Let a unitary operator $R$ is defined to act on the first register $|x\rangle_1$ with $n$-qubit and auxiliary second register $|0\rangle_2$ with 1-qubit to rotate the second register when $f(x)$ is bounded between 0 and 1 as follows:

$$R|x\rangle_1|0\rangle_2 = \sqrt{f(x)}|x\rangle_1|1\rangle_2 + \sqrt{1-f(x)}|x\rangle_1|0\rangle_2. \tag{9}$$

Let unitary operation $P$ be an operation for building a superposition of all $x$ states from $|0\rangle_1$ state as follows:

$$P|0\rangle_1 = \sum_{x=0}^{2^n-1} \sqrt{p(x)}|x\rangle_1. \tag{10}$$

Then, by applying operators $P$ and $R$ to the initialized quantum registers, the expected value appears as a probability of measuring $|1\rangle_2$ in the second register as follows:

$$R(P \otimes I_2)|0\rangle_1|0\rangle_2 = \sum_{x=0}^{2^n-1} \sqrt{f(x)}\sqrt{p(x)}\,|x\rangle_1|1\rangle_2 + \sum_{x=0}^{2^n-1} \sqrt{1-f(x)}\sqrt{p(x)}\,|x\rangle_1|0\rangle_2, \tag{11}$$

where $I_2$ denotes an identity operator on the second register. Therefore, the expected value can be estimated by amplitude estimation in the state where the second register is $|1\rangle_2$. The total number of times of calling the operation $A=R(P \otimes I_2)$ and $A^{-1}$ require

$O(1/\epsilon)$ in the quantum amplitude estimation, which squares faster than a classical Monte Carlo method [6].

### 3.2 Quantum circuit

Pi can be estimated from the area ratio of a square and a quadrant inscribed in a square whose radius is equal to the side length of the square, which is a popular algorithm. By preparing $n$-qubit registers to build superposition of integers representing binary numbers on $x$ and $y$ axes, respectively, a total of $2^{2n}$ points in a square with a side length of $2^n$ can be sampled as shown in Fig. 9. The function that determines whether the sampled point $(x, y)$ is inside the quadrant or not is as follows:

$$f(x,y) = \begin{cases} 1 \text{ if } x^2 + y^2 < 2^{2n} \\ 0 \text{ if } x^2 + y^2 \geq 2^{2n} \end{cases}. \tag{12}$$

Herein, pi can be estimated as follows:

$$\frac{\pi}{4} = \sum_{y=0}^{2^n-1} \sum_{x=0}^{2^n-1} f(x,y) p(x,y), \tag{13}$$

where $p(x, y)$ denotes the probability of sampling point $(x, y)$.

In the quantum circuit, three quantum registers are used to estimate pi: the first quantum register with $n$-qubit for sampling along the $x$-axis, the second quantum register with $n$-qubit for sampling along the $y$-axis, and the third quantum register with 1-qubit for amplitude estimation. Thus, the operator $P$ and $R$ in this algorithm act on two and three quantum registers, respectively, as follows:

$$R(P \otimes I_3)|0\rangle_1|0\rangle_2|0\rangle_3 = R \sum_{y=0}^{2^n-1} \sum_{x=0}^{2^n-1} \sqrt{p(x,y)} \, |x\rangle_1|y\rangle_2|0\rangle_3$$

$$= \sum_{y=0}^{2^n-1} \sum_{x=0}^{2^n-1} \sqrt{f(x,y)} \sqrt{p(x,y)} \, |x\rangle_1|y\rangle_2|1\rangle_3$$

$$+ \sum_{y=0}^{2^n-1} \sum_{x=0}^{2^n-1} \sqrt{1-f(x,y)} \sqrt{p(x,y)} \, |x\rangle_1|y\rangle_2|0\rangle_3, \tag{14}$$

where $I_3$ denotes an identify operation on the third quantum register. Hence, pi can be estimated by using the quantum amplitude estimation method when the third register is $|1\rangle_3$.

The operation *P* builds superposition of *x* and *y* with the same probability, and it is implemented by applying Hadamard gates on each qubit of the first and second registers.

The operation *R* rotates the third register according to the state of the first and second registers and the amplitude become $\sqrt{f(x,y)}$ when the state is $|1\rangle_3$. Further, this rotated state can be represented as follows since *f*(*x*, *y*) takes only 0 or 1:

$$\sqrt{f(x,y)}|1\rangle_3 + \sqrt{1-f(x,y)}|0\rangle_3 = |f(x,y)\rangle_3. \tag{15}$$

The operation *R* implemented in a quantum circuit is shown in Fig. 10. This circuit requires an ancillary quantum register with 2*n*-qubit initialized to $|0\rangle_A$. First, the ancillary quantum register is combined with the third quantum register so that the third register is the most significant bit, which is referred to as 3A register. Then, squared values of *x* and *y* are added in the 3A register by quantum squarer proposed in section 2.2. When the 3A register is split into original 1 and 2*n* qubits, the state is represented as $|\overline{f(x,y)}\rangle_3 |x^2 + y^2 \bmod 2^{2n}\rangle_A$ because the third quantum register gets $|1\rangle_3$ only if $x^2 + y^2$ is more than $2^{2n}$. By attaching the X gate to the third quantum register, the state becomes $|f(x,y)\rangle_3$. The ancillary quantum register can be undone by performing inverse quantum squarer operation because quantum squarer is a unitary operation. These operations are summarized as follows:

$|x\rangle_1 |y\rangle_2 |0\rangle_3 |0\rangle_A$
$\rightarrow |x\rangle_1 |y\rangle_2 |0\rangle_{3A}$            Combining 3rd and anciilary regsiter
$\rightarrow |x\rangle_1 |y\rangle_2 |x^2 + y^2 \bmod 2^{2n+1}\rangle_{3A}$     Squaring *x* and *y*
$\rightarrow |x\rangle_1 |y\rangle_2 |\overline{f(x,y)}\rangle_3 |x^2 + y^2 \bmod 2^{2n}\rangle_A$    Spliting 3A regsiter
$\rightarrow |x\rangle_1 |y\rangle_2 |f(x,y)\rangle_3 |x^2 + y^2 \bmod 2^{2n}\rangle_A$    NOT opeartion on 3rd regsiter
$\rightarrow |x\rangle_1 |y\rangle_2 |f(x,y)\rangle_3 |0\rangle_A$            Undoing ancillary register

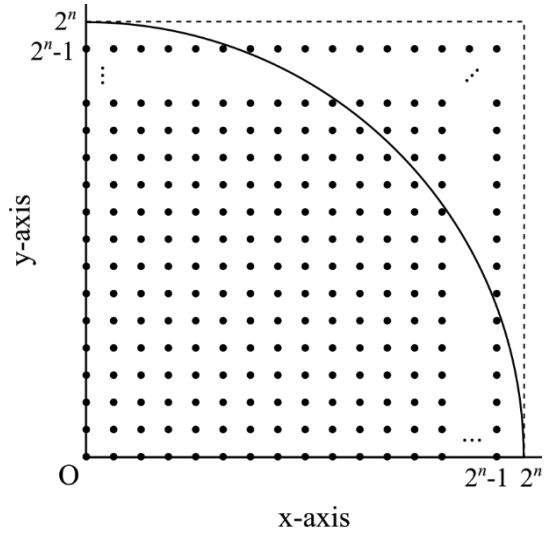

**Fig. 8** Sampling points in a squarer with a side length of $2^n$ and an inscribed quadrant

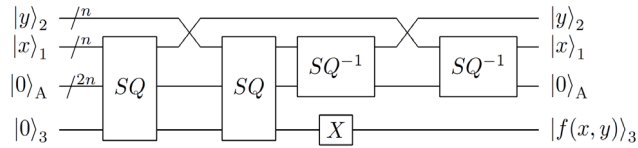

**Fig. 9** Quantum circuit of operation $R$ to estimate pi. In the circuit, $SQ$ and $SQ^{-1}$ refer to the quantum squarer circuit and its inverse circuit, respectively

## *3.3 Simulation*

We implemented the quantum circuit to estimate pi in Qiskit 0.19.2 [21, 22] and performed the simulation using its quantum computer simulator. The QFT squarer proposed in section 2.2.2 and quantum amplitude estimation method using maximum likelihood [5] (to avoid increasing the ancillary qubits) were used in the circuit. The total number of qubits including ancillary qubits is $4n + 1$ while the total number of sampling is $2^{2n}$.

The computational complexity of the quantum amplitude estimation method is defined by the total number of calling the operation $A=R(P \otimes I_3)$ and $A^{-1}$, and that of using maximum likelihood is $O((1/\epsilon_A)\ln(1/\epsilon_A))$, where $\epsilon_A$ is the estimation error of the quantum amplitude estimation. The method estimates the quantum amplitude by applying an operator including the $A$ and $A^{-1}$ for $m_k$ times and by using maximum likelihood on multiple circuit sets with different $k$. The simulations were performed for two patterns: $k_{max}$ = 1 and 5, which are circuit sets with $k$ = 0 to $k_{max}$ with $m_0$ = 0 and $m_k$ = $2^{k-1}$. For the 100 shots per circuit, the computational complexity of the former is 400 and $\epsilon_A$ is $O(10^{-2})$, while it is 6,800 and $O(10^{-3})$ for the later.

During the simulation, pi was estimated 100 times for each $n$ while $n$ was varied from 2 to 6 and the averaged values and standard deviations of pi were recorded. Additionally, the expected values obtained by using systematic samplings were calculated by performing classical calculations.

### *3.4 Results and discussion*

A plot of the average estimated values and standard deviations obtained from the quantum simulations and the expected value obtained through classical calculations is shown in Fig. 11. The results of quantum simulations are in line with those obtained by performing classical calculations. However, errors $\epsilon_S$ from the sampling numbers chosen have to be considered to estimate pi. While the number of samples and the computational complexity are the same in classical Monte Carlo method, they are different in the quantum circuit. The $\epsilon_S$ is $O(1/\sqrt[d]{N})$, where $d$ denotes the dimension of the problem and $N$ denotes the total number of samples. In the case of pi estimation, $\epsilon_S$ is $O(1/2^n)$ and the total estimation error is $\epsilon_S + \epsilon_A$. The computational complexity can be decreased by using original quantum amplitude estimation of $O(1/\epsilon_A)$ [5], the number of qubits used for sampling in each dimension is expected to require $O(\log_2(1/\epsilon_A))$ or more to produce sufficient accuracy.

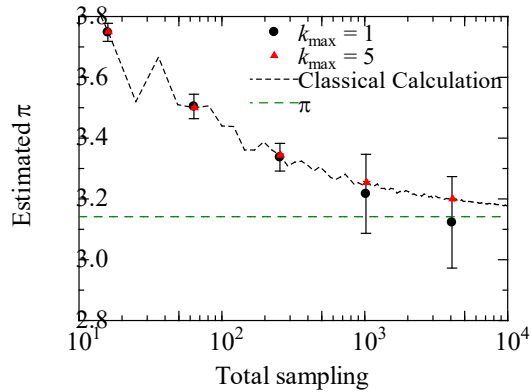

**Fig. 10** Estimated pi from the quantum simulations and classical calculation

## 4. Conclusion

In this study, quantum circuits for multiplying, squaring, and estimating pi are proposed. The proposed quantum multiplier can be used to calculate different digits of binary numbers by using known quantum adder. In the case of multiplication of two $n$-qubit binary numbers using quantum adder with gate size of $O(n)$, the multiplier requires $O(n^2)$ gate and at least $n$ ancillary qubits while known QFT multiplier requires $O(n^3)$ gate without any ancillary qubits.

Two types of quantum multiplier based quantum squarers are also proposed, which reduce input quantum register while requiring the same order of gate size and ancillary qubits as the quantum multiplier.

A quantum algorithm to estimate pi based on Monte Carlo method and quantum amplitude estimation was implemented and demonstrated using a quantum computer simulator. This circuit required $4n + 1$ qubits for a total of $2^{2n}$ sampling size. The estimated values contain $\epsilon_S$ and $\epsilon_A$ errors, which are caused by the sampling number chosen and computational complexity, respectively. To obtain higher accuracy, the circuit requires an adequate number of qubits for sampling that corresponds to the computational complexity.

## Acknowledgments

The author thanks Prof. Sougato Bose and Dr. Hidekazu Kurebayashi from University College London for their support and advice.